# Local structural ordering determines the mechanical damage tolerance of amorphous grain boundary complexions

Pulkit Garg [a], Timothy J. Rupert [a,b,*]

[a] Department of Materials Science and Engineering, University of California, Irvine, CA 92697, USA

[b] Department of Mechanical and Aerospace Engineering, University of California, Irvine, CA 92697, USA

* To whom correspondence should be addressed: trupert@uci.edu

Amorphous grain boundary complexions act as toughening features within a microstructure because they can absorb dislocations more efficiently than traditional grain boundaries. This toughening effect should be a strong function of the local internal structure of the complexion, which has recently been shown to be determined by grain boundary crystallography. To test this hypothesis, molecular dynamics are used here to simulate dislocation absorption and damage nucleation for complexions with different distributions of structural short-range order. The complexion with a more disordered structure away from the dislocation absorption site is actually found to better resist crack nucleation, as damage tolerance requires delocalized deformation and the operation of shear-transformation zones through the complexion thickness. The more damage tolerant complexion accommodates plastic strain efficiently within the entire complexion, providing the key mechanistic insight that local patterning and asymmetry of structural short-range order controls the toughening effect of amorphous complexions.

Solute segregation can lead to structural transitions between different types of *grain boundary complexions*, or interfacial phases in thermodynamic equilibrium with their abutting grains [1, 2]. Nanoscale film grain boundary complexions have a stable, finite thickness and can be structurally ordered or disordered depending on grain boundary crystallography, temperature, and chemistry [3, 4]. While ordered complexions are often reported to be brittle and lead to premature intergranular failure in generally ductile metals [5, 6], structurally disordered or amorphous complexions can enhance the mechanical damage resistance of nanocrystalline alloys [7, 8]. Amorphous complexions have been shown to simultaneously improve the strength and ductility of nanocrystalline Cu-Zr alloys, while also increasing the thermal stability [9, 10], as compared to nanocrystalline Cu [11, 12]. Thicker amorphous complexions can lead to further improvements in the ductility of nanocrystalline alloys as these features can accommodate more dislocations before damage nucleation [13, 14].

The properties of complexions are connected to their internal structure [15], and in turn, play an important role in determining the bulk properties of polycrystalline materials [16]. Amorphous complexions can be characterized by the structural short-range order (SRO), defined as the reoccurrence of the same type of atomic arrangement within the first coordination shell [17]. For example, amorphous complexions can be separated into two structurally distinct regions: (1) the complexion interior with a fully amorphous structure and (2) the amorphous-crystalline transition regions (ACTRs), which connect the interior with neighboring crystalline grains [18]. Our recent work shows that the structure of amorphous complexions is controlled by the incompatibility or misfit between the neighboring grains with the more strained neighboring crystal leads to a more disordered ACTR [19].

Structural SRO in bulk amorphous materials has been directly connected to mechanical properties [20, 21], which can serve as a reference point for amorphous complexions. For example, in Cu-Zr metallic glasses, geometrically favorable icosahedral polyhedra have been shown to provide resistance to plastic flow, leading to increased strength and ductility [22, 23]. These geometrically favorable structures are contrasted by geometrically unfavorable motifs (GUMs) which fill the remaining material and connect the foundational icosahedral polyhedra in bulk metallic glasses [24]. GUMs have highly disordered atomic arrangements that form instability prone soft spots [25, 26], which upon deformation are responsible for the localized operation of shear transformation zones (STZs) that leads to the fracture of metallic glasses [27, 28]. Thus, the importance of local atomic structure for the deformation behavior of bulk amorphous metals is established, yet such features are not understood for amorphous complexions. We hypothesize that the dislocation absorption capacity and toughening ability of amorphous complexions should be controlled by the atomic details of the internal complexion structure.

In this study, the connection between structural SRO and the damage tolerance of amorphous complexions is probed using atomistic simulations. Two Cu-Zr bicrystal samples with different combinations of incompatible neighboring grains were created, resulting in structurally distinct amorphous complexions. Importantly, the dislocation absorption capacity of one complexion is found to be twice as high as the other. The damage nucleation process is delayed when plasticity can be shared through the complexion thickness, with GUMs found to accommodate the majority of the plastic strain. This work establishes the first structure-property relationship for amorphous complexions by demonstrating the impact of structural SRO on damage tolerance.

Amorphous complexion models were created with hybrid Monte Carlo/molecular dynamics (MD) simulations using the Large-scale Atomic/Molecular Massively Parallel Simulator (LAMMPS) code [29], with a 1 fs integration time step for all MD simulations and an embedded-atom method potential for Cu-Zr [30]. Two starting Cu bicrystal samples were chosen with different crystallographies. The X-axis of the first grain, Grain A, is oriented along the [110] direction in both bicrystal samples, whereas the X-axis of the second neighboring grain is oriented along the [1$\bar{1}$1] direction (Grain B) in Complexion 1 and along the [15$\bar{2}$] direction (Grain C) in Complexion 2 (Fig. 1(a)). The simulation cell is approximately 115 nm long (X-direction), 31 nm tall (Y-direction), and 10 nm thick (Z-direction), containing ~3,000,000 atoms, with periodic boundary conditions in all directions. The bicrystal samples were first relaxed at 1000 K using a Nose-Hoover thermo/barostat at zero pressure, followed by doping with 2 at.% Zr and further equilibration, with additional details provided in Ref. [19]. The complexion samples were quenched from 1000 K to 10 K at a cooling rate of $10^{13}$ K/s while maintaining zero pressure in all directions, followed by a hold at 10 K for 100 ps under zero pressure.

A positive hydrostatic stress was created to provide a driving force for crack/damage nucleation by applying an elastic tensile strain of 4% in the X-direction while not allowing Poisson contraction in the other directions. The samples were then held for another 200 ps under the canonical ensemble after the tensile pre-strain step. An artificial dislocation source in the center of the samples was operated by gradually displacing two layers of atoms relative to each other at a constant speed to generate multiple dislocation pairs, following the procedure introduced by Pan and Rupert [31]. Simultaneously, a shear deformation under the canonical ensemble was applied at an engineering shear strain rate of $10^8$ s$^{-1}$, with the two bottom layers of atoms held fixed in the Y-direction to limit rigid body grain rotation. The dislocations nucleated from the dislocation source upon deformation were identical in both of the complexion samples as the center grain (Grain A), the dislocation source, and the simulation cell deformation are all the same for the two samples. Shockley partial dislocations with $\vec{b} = a/6 \langle 112 \rangle$ are created and then absorbed in all cases. For both bicrystals, five thermodynamically equivalent configurations differing slightly by subtle thermal vibrations were examined to allow for improved statistics. The local structure of each atom was characterized using the Voronoi tessellation method [32, 33] with the index notation <$n_3$, $n_4$, $n_5$, $n_6$>, where $n_i$ stands for the number of Voronoi polyhedron faces with $i$ edges. The density of specific polyhedra types then provides a metric to describe SRO variations within the complexions. Since Cu is the primary elemental species in the alloys studied here, our analysis was restricted to the structural SRO motifs observed around Cu atoms. Structural analysis and visualization were performed using the open-source visualization tool OVITO [34].

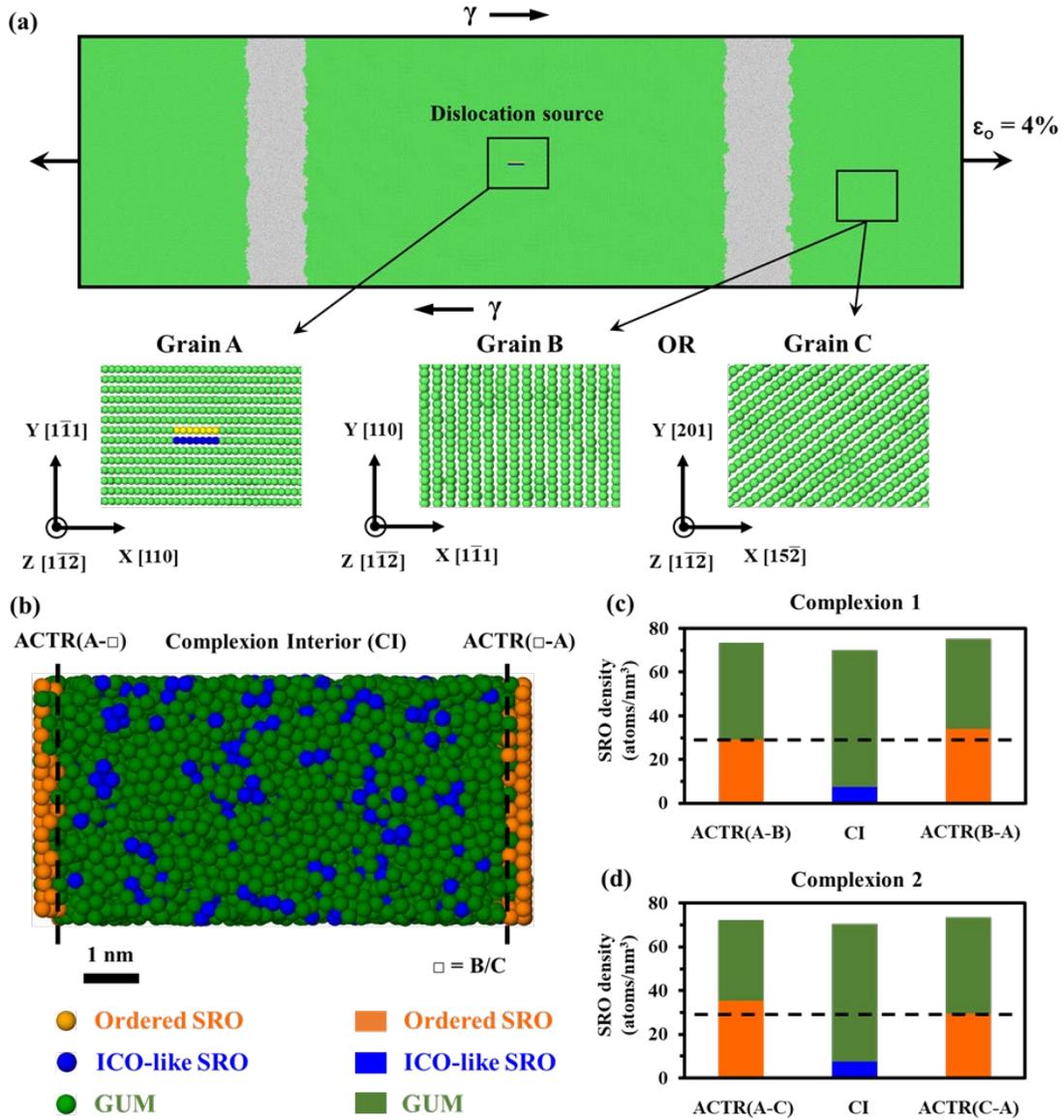

Fig. 1. (a) A Cu-Zr bicrystal sample containing amorphous grain boundary complexions, where relative grain orientations can be varied. (b) Image of a representative amorphous complexion separated into different regions based on structural SRO and the nearest grain. Ordered SRO is most commonly observed in ACTRs, whose edges are marked by dashed black lines, while more disordered SRO dominates the complexion interior (CI). The density of SRO types in different regions of (c) Complexion 1 and (d) Complexion 2. The black dashed lines mark the ordered SRO density in the more disordered ACTR, which occurs on the inside of Complexion 1 and the outside of Complexion 2. Therefore, ACTR(A-B) is less ordered than ACTR(B-A) in Complexion 1, while ACTR(A-C) is more ordered than ACTR(C-A) in Complexion 2.

Fig. 1(b) shows an undeformed amorphous complexion separated into different regions with distinct SRO patterns. The local structure of the ACTRs is notable for the presence of <0, 10, 2, 0> and <0, 8, 4, 0> polyhedra, which are similar to the perfect FCC polyhedra <0, 12, 0, 0> and therefore referred to as *ordered SRO* polyhedra (represented by orange atoms). The central region of the complexion corresponds to the complexion interior (CI) and is dominated by <0, 0, 12, 0> and <0, 2, 10, 0> polyhedra, which are similar to the amorphous structure of bulk Cu-Zr glasses and together referred to as Icosahedral-like (*ICO-like*) motifs (represented by blue atoms). The remaining CI region is filled with a variety of highly disordered GUMs (represented by green atoms). The ACTR edge is defined at the location where the ordered polyhedra become more common than the ICO-like polyhedra, resulting in ACTR thicknesses of ~0.6 nm for the complexions studied in this work. Further, different ACTRs can be distinguished based on their nearest grain such that the ACTR next to Grain A is referred to as ACTR (A-B) and that adjacent to Grain B is named ACTR(B-A) in Complexion 1. Similarly, the ACTR next to Grain A is referred to as ACTR(A-C) while the one next to Grain C is called ACTR(C-A) in Complexion 2.

The incoming dislocations interact with ACTR(A-B) and ACTR(A-C) in Complexion 1 and Complexion 2, respectively. Figs. 1(c) and (d) show the density of ordered, ICO-like, and GUM SRO types in the different regions of the complexions. The density of ICO-like polyhedra and GUMs in the two CI are nearly identical and also similar to the local structural order in the bulk amorphous phase [35]. However, an asymmetry is observed in the ordered SRO density of the ACTRs, where one side is more ordered than the other. In Complexion 1, the incoming dislocations will be absorbed at the less ordered ACTR(A-B), while the dislocations will be absorbed at the more ordered ACTR(A-C) in Complexion 2. Both the complexions have a thickness of ~8 nm, comparable to prior experimental observations in Cu- [9] and Al-rich [36] alloys, and similar SRO density in different regions of the two complexions. The primary difference between the two is the reversal in the location of the more ordered ACTR, which appears next to the exterior grain in Complexion 1 and next to the interior grain (i.e., where the dislocations are absorbed) in Complexion 2. The systematic difference between the complexion structures provides a straightforward method to test our hypothesis that local structure determines damage tolerance.

Fig. 2(a) and (b) show the initial damage nucleation event at ACTR(A-B) and ACTR(A-C) in Complexion 1 and Complexion 2, respectively. This point is identified when the total void/crack volume is greater than 0.5 nm$^3$, equivalent to the size of a spherical crack with a diameter of ~1 nm. The damage resistance of the two complexions differs significantly, as damage nucleation occurs in Complexion 1 at a shear strain of 6.5% whereas, in Complexion 2 damage nucleated at a shear strain of 8.9%. More important than the critical strain value is the fact that the dislocation absorption capacity of Complexion 2 is two times higher than that of Complexion 1, as this is the physically relevant event associated with plasticity and damage nucleation. Complexion 2 accommodated three full dislocations before damage nucleation, whereas Complexion 1 absorbed only one full dislocation and half of the second dislocation (the leading Shockley partial) before damage nucleation. The ordered grain boundaries from which the two complexion samples were created have identical damage tolerances, with both absorbing only one partial dislocation before damage nucleation, as shown in Fig. S1 in the Supplemental Materials. Therefore, it is clear that subtle differences in the local structure of complexions arising from variations in grain boundary crystallography can significantly improve their damage tolerance.

Fig. 2(c) shows the damage site at ACTR(A-B) along with the surface atoms around this site, which primarily consist of GUMs and a smaller population of ordered and ICO-like SRO. To understand the effect of dislocation absorption and damage on complexion SRO, Figs. 2(d) and (e) show the density of ordered SRO and GUMs as a function of applied shear strain in the ACTRs that interact with incoming dislocations in each complexion. The ordered SRO density in both complexions remains constant for a period and then decreases as shear strain increases towards the damage nucleation point (gray dashed line). The opposite trend is observed in the GUM density, which increases near the damage nucleation point. Thus, several atoms with ordered SRO are transforming to GUMs as damage begins, with an example of one such structural transition shown in Fig. 2(f). The conversion of geometrically favorable motifs to GUMs has been widely observed during the deformation of bulk metallic glasses, which promotes the activation of STZs leading to shear banding and fracture [37, 38], with our results showing the first such observation made in amorphous complexions.

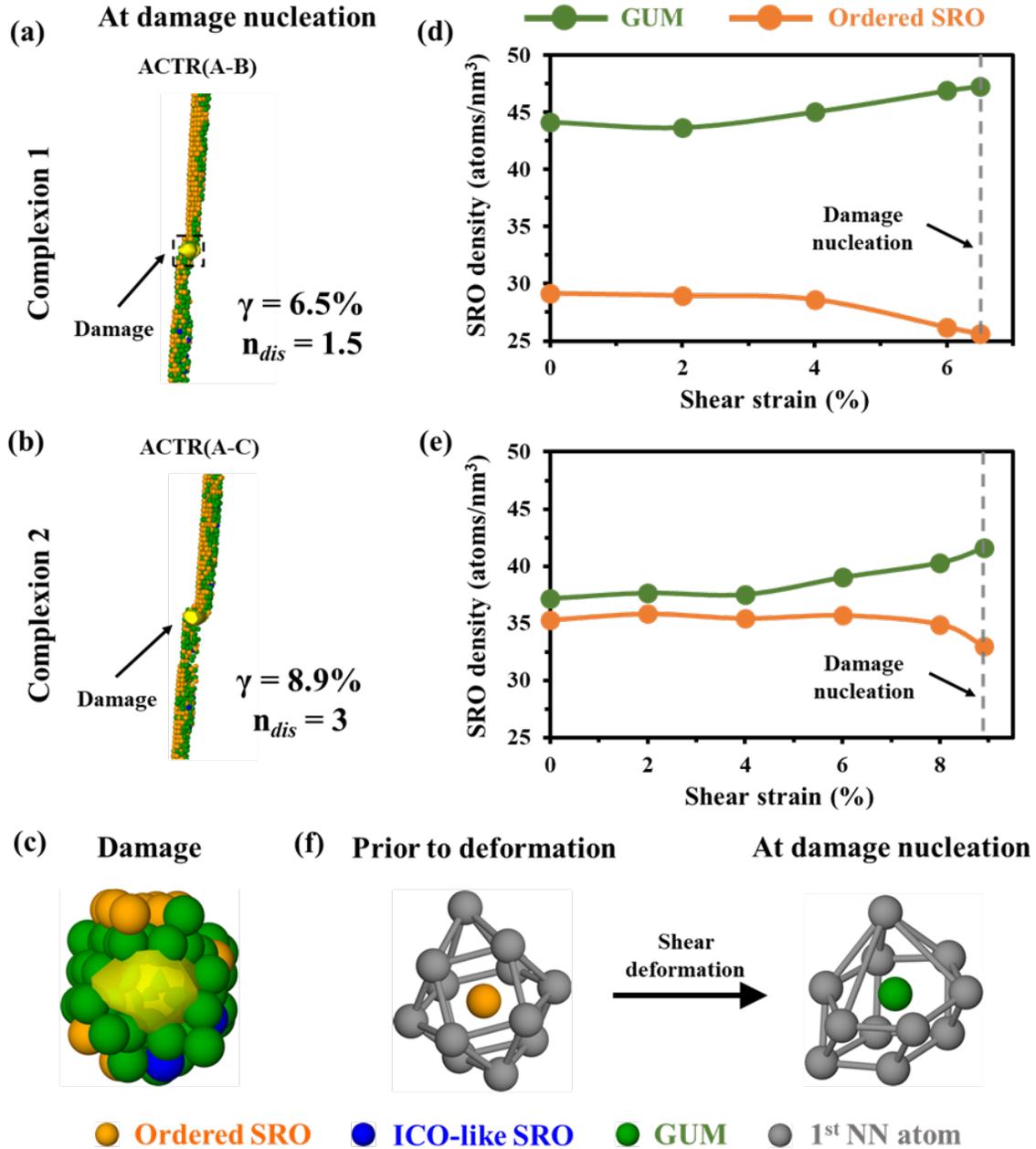

Fig. 2. Damage nucleation at the ACTRs where dislocations are absorbed in (a) Complexion 1 and (b) Complexion 2, with the shear strain and the number of dislocations absorbed at damage nucleation noted. (c) A damaged region with the surface atoms colored according to their structural SRO. The density of ordered SRO and GUMs as a function of applied shear strain in the ACTRs where dislocations are absorbed in (d) Complexion 1 and (e) Complexion 2. (f) An atom that ends up being along the damage surface undergoing a structural transformation from ordered motif to GUM upon deformation.

To understand why one complexion is twice as damage tolerant than the other, the distribution of plasticity within the complexions is next investigated using the non-affine squared displacement (NASD), a deformation descriptor associated with the activation of STZs in metallic glasses [39, 40]. NASD is calculated by measuring the non-elastic strain experienced by an atom with reference to the undeformed sample [41]. An atom is considered to undergo plastic deformation when its NASD exceeds the square of the distance of the second nearest neighbor shell (NASD > 20 Å$^2$, for this alloy system) [42]. The STZs, therefore, are simply clusters of atoms with high NASD values [43, 44]. To allow for a rigorous comparison, Fig. 3 presents data from the complexions on the right side of the simulation cells. Identical observations were made from the complexions on the left side and are presented in Fig. S5 in the Supplemental Materials. Figs. 3(a) and (b) show the snapshots of the NASD distribution in the two amorphous complexions before damage (one fewer Shockley partial has been absorbed) and at damage nucleation. The distribution of plasticity is quantified by measuring the density of STZs within a 1 nm thick strip normal to the dislocation interaction region, as shown by the gray boxes in Figs. 3(a) and (b). Fig. 3(c) and (d) present the spatial distribution of STZ density. In Complexion 1, very limited STZs operate before damage which are primarily concentrated at the dislocation absorption location within ACTR(A-B), and only a slight increase is observed in the amorphous interior of complexion upon damage nucleation. In contrast, STZs are observed not only around the dislocation absorption region in ACTR(A-C) but also within the CI and along ACTR(C-A) on the other side of Complexion 2 both before damage and at damage nucleation. These dramatic differences in the deformability of the two complexions demonstrates the importance of subtle variations in the local structure. The localization of STZs is shown to be responsible for the formation of shear bands and catastrophic failure in metallic glasses [45, 46]. Amorphous complexions are damage tolerant because of their shared plasticity, and Complexion 2 clearly shares this plastic deformation into a larger region. Complexion 2 has less order at the opposite ACTR and is, therefore, able to recruit that region to help relieve the strain heterogeneities from the incoming dislocations, delaying damage nucleation. It is more damage tolerant because it more efficiently uses the entire amorphous structure to distribute plasticity within the complexion. Complexion 2 survives to much larger applied strains, at which point dislocations can be observed to nucleate into the neighboring grains, as shown in Fig. S4 in the Supplemental Materials. The sharing of plastic deformation with the next grain may provide an additional pathway to toughening.

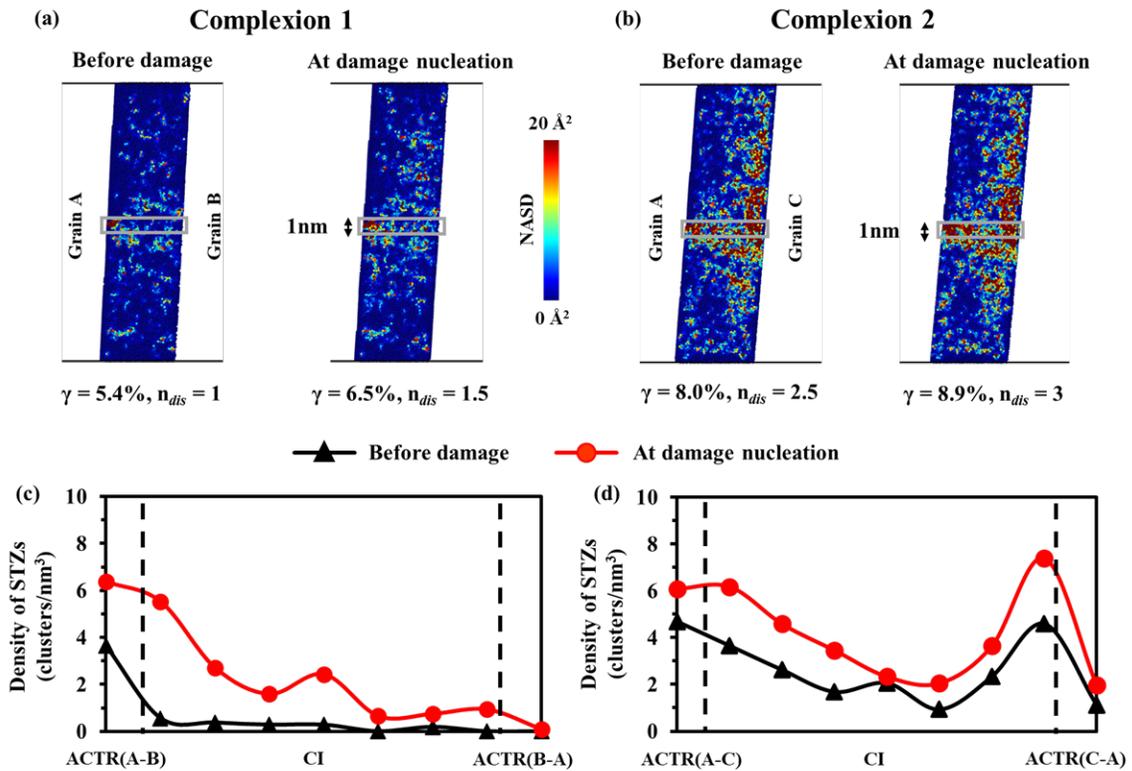

Fig. 3. The distribution of NASD before damage and at damage nucleation in (a) Complexion 1 and (b) Complexion 2. Atoms with NASD > 20 Å$^2$ have deformed plastically through STZ operation, with the density of STZs quantified within a 1 nm thick strip normal to the dislocation absorption site. The spatial distributions

**of STZ density are plotted for (c) Complexion 1 and (d) Complexion 2 with the edges of the ACTRs marked by the dashed black lines.**

Finally, one must understand why one complexion deforms in a more distributed manner than the other. Multiple structural parameters such as free volume [47], configurational potential energy [48], local entropy [49], and atomic-level stresses [50] have been used to build structure-property relationships and explain plasticity in metallic glasses, yet these all have limitations [38]. Free volume is the most commonly used parameter to characterize the structure of amorphous materials [51], although it was originally developed for liquid and gaseous systems and lacks accuracy for solid materials [52]. Ding et al. [38] proposed a new structural parameter, *flexibility volume* ($v_{flex}$), which takes into account both atomic volume and interactions between neighbors via atomic vibrations. This parameter quantitively predicted the mechanical properties of metallic glasses, with local regions of higher flexibility volume resulting in a more ductile material [53]. Flexibility volume is an indicator of the propensity for plasticity in glasses as regions with high flexibility volume have a higher tendency of STZ activation and is defined as the product of atomic vibrational mean squared displacement (MSD) and the average atomic spacing, as explained in Refs. [53, 54]. The undamaged states of the two complexion samples, without any shear deformation, were equilibrated under a microcanonical ensemble (NVE) at 300 K to calculate the vibrational MSD over a short period of time, while local atomic spacing is determined by taking a cube root of the local atomic volume as measured by the Voronoi analysis method in OVITO. Figs. 4(a) and (b) show the distribution of the flexibility volume within Complexion 1 and Complexion 2, respectively. In Complexion 1, a few randomly distributed regions exhibit high $v_{flex}$. In contrast, in Complexion 2 there is a large population of interconnected high $v_{flex}$ regions along the side of the complexion opposite from the dislocation absorption site, near ACTR(C-A).

These regions are primed to easily flow, enabling the large density of STZs, distributed plasticity, and improved damage tolerance shown in Fig. 3(b) for Complexion 2.

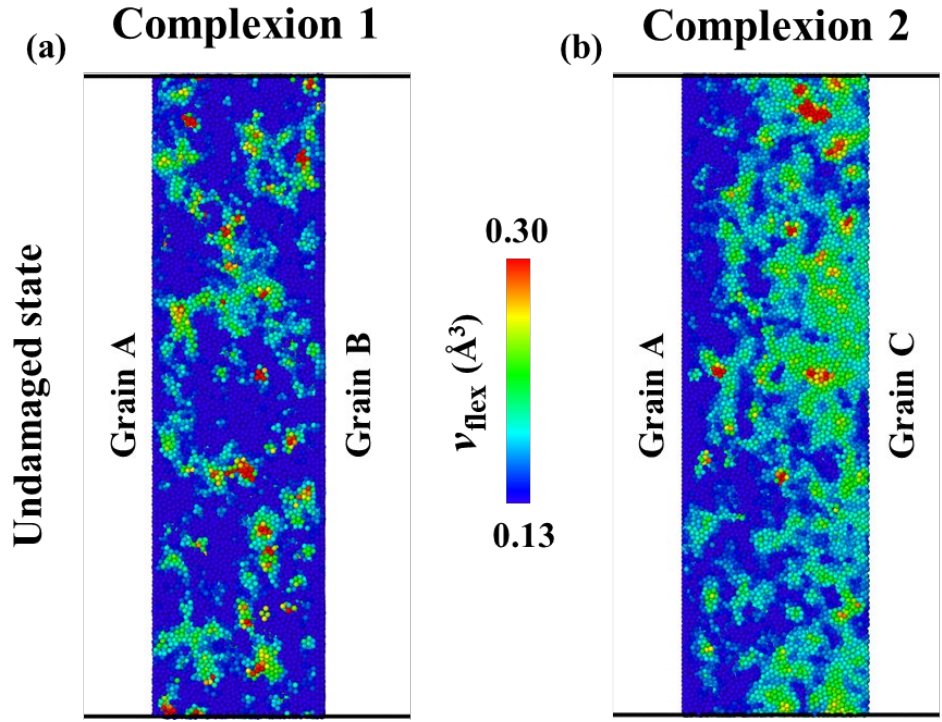

**Fig. 4. Spatial distribution of the atomic flexibility volume ($v_{flex}$) in (a) Complexion 1 and (b) Complexion 2 in their undamaged states. Limited high $v_{flex}$ regions are observed in Complexion 1 while a large population of high $v_{flex}$ is present in Complexion 2 along the side further away from the dislocation absorption site, enabling distributed plasticity.**

In summary, the damage tolerance of amorphous complexions was investigated with MD simulations, with a focus on understanding how spatial variations in SRO affect performance. Ordered SRO transforms to highly disordered GUMs upon shear deformation, leading to damage nucleation near the dislocation absorption site. One of the complexions was found to be twice as damage tolerant as the other because it could accommodate the plastic strain from incoming dislocations more efficiently by sharing deformation through the complexion thickness. The effect of SRO patterning on the damage tolerance of complexions is comparable to grain boundary deformation in nanocrystalline alloys which is significantly affected by variations in the atomic structure of interfaces [55]. The structural order/disorder and the corresponding propensity for activation of STZs in amorphous complexions were shown to be related to the flexibility volume structural parameter. A large number of easily deformed regions through the complexion thickness is beneficial for damage tolerance because it allows for distributed plasticity. Thus, the damage tolerance of an amorphous complexion is indeed directly determined by its local atomic structure, with this work providing a pathway for the design of strong and tough interfaces.

**Acknowledgments**

This research was supported by the U.S. Department of Energy, Office of Science, Basic Energy Sciences, under Award No. DE-SC0021224. Structural analysis and atomic visualization were performed with software supported by the National Science Foundation Materials Research Science and Engineering Center program through the UC Irvine Center for Complex and Active Materials (DMR-2011967).

**References**

[1] S.J. Dillon, M. Tang, W.C. Carter, M.P. Harmer, Complexion: a new concept for kinetic engineering in materials science, Acta Materialia 55(18) (2007) 6208-6218.
[2] P.R. Cantwell, M. Tang, S.J. Dillon, J. Luo, G.S. Rohrer, M.P. Harmer, Grain boundary complexions, Acta Materialia 62 (2014) 1-48.
[3] S.J. Dillon, G.S. Rohrer, Mechanism for the development of anisotropic grain boundary character distributions during normal grain growth, Acta Materialia 57(1) (2009) 1-7.
[4] W.-R. Jian, Z. Xie, S. Xu, Y. Su, X. Yao, I.J. Beyerlein, Effects of lattice distortion and chemical short-range order on the mechanisms of deformation in medium entropy alloy CoCrNi, Acta Materialia 199 (2020) 352-369.
[5] W. Sigle, G. Richter, M. Rühle, S. Schmidt, Insight into the atomic-scale mechanism of liquid metal embrittlement, Applied Physics Letters 89(12) (2006) 121911.
[6] J. Luo, H. Cheng, K.M. Asl, C.J. Kiely, M.P. Harmer, The role of a bilayer interfacial phase on liquid metal embrittlement, Science 333(6050) (2011) 1730-1733.
[7] Y.M. Wang, A.V. Hamza, T.W.B. Jr., Incipient plasticity in metallic glass modulated nanolaminates, Applied Physics Letters 91(6) (2007) 061924.
[8] A. Khalajhedayati, Z. Pan, T.J. Rupert, Manipulating the interfacial structure of nanomaterials to achieve a unique combination of strength and ductility, Nature Communications 7(1) (2016) 1-8.
[9] C.M. Grigorian, T.J. Rupert, Thick amorphous complexion formation and extreme thermal stability in ternary nanocrystalline Cu-Zr-Hf alloys, Acta Materialia 179 (2019) 172-182.
[10] A. Khalajhedayati, T.J. Rupert, High-temperature stability and grain boundary complexion formation in a nanocrystalline Cu-Zr alloy, JOM 67(12) (2015) 2788-2801.
[11] J.L. Wardini, C.M. Grigorian, T.J. Rupert, Amorphous complexions alter the tensile failure of nanocrystalline Cu-Zr alloys, Materialia (2021) 101134.
[12] Y. Wang, J. Li, A.V. Hamza, T.W. Barbee, Ductile crystalline–amorphous nanolaminates, Proceedings of the National Academy of Sciences 104(27) (2007) 11155-11160.
[13] Z. Pan, T.J. Rupert, Amorphous intergranular films as toughening structural features, Acta Materialia 89 (2015) 205-214.
[14] S. Pal, K. Vijay Reddy, C. Deng, On the role of Cu-Zr amorphous intergranular films on crack growth retardation in nanocrystalline Cu during monotonic and cyclic loading conditions, Computational Materials Science 169 (2019) 109122.
[15] A. Sutton, R. Balluffi, Interfaces in crystalline materials, Oxford, Oxford Science Publications. (1995).
[16] Y. Mishin, M. Asta, J. Li, Atomistic modeling of interfaces and their impact on microstructure and properties, Acta Materialia 58(4) (2010) 1117-1151.
[17] Y. Zhang, F. Zhang, C. Wang, M. Mendelev, M. Kramer, K. Ho, Cooling rates dependence of medium-range order development in C u 64. 5 Z r 35. 5 metallic glass, Physical Review B 91(6) (2015) 064105.
[18] Z. Pan, T.J. Rupert, Spatial variation of short-range order in amorphous intergranular complexions, Computational Materials Science 131 (2017) 62-68.
[19] P. Garg, T.J. Rupert, Grain incompatibility determines the local structure of amorphous grain boundary complexions, Acta Materialia 244 (2023) 118599.
[20] Y. Cheng, H. Sheng, E. Ma, Relationship between structure, dynamics, and mechanical properties in metallic glass-forming alloys, Physical Review B 78(1) (2008) 014207.

[21] Y. Cheng, E. Ma, Intrinsic shear strength of metallic glass, Acta Materialia 59(4) (2011) 1800-1807.
[22] Y. Cheng, A.J. Cao, H. Sheng, E. Ma, Local order influences initiation of plastic flow in metallic glass: Effects of alloy composition and sample cooling history, Acta Materialia 56(18) (2008) 5263-5275.
[23] K.-W. Park, J.-i. Jang, M. Wakeda, Y. Shibutani, J.-C. Lee, Atomic packing density and its influence on the properties of Cu–Zr amorphous alloys, Scr. Mater. 57(9) (2007) 805-808.
[24] D. Şopu, F. Moitzi, N. Mousseau, J. Eckert, An atomic-level perspective of shear band formation and interaction in monolithic metallic glasses, Applied Materials Today 21 (2020) 100828.
[25] J. Ding, Y.Q. Cheng, E. Ma, Quantitative measure of local solidity/liquidity in metallic glasses, Acta Materialia 61(12) (2013) 4474-4480.
[26] J. Ding, S. Patinet, M.L. Falk, Y. Cheng, E. Ma, Soft spots and their structural signature in a metallic glass, Proceedings of the National Academy of Sciences 111(39) (2014) 14052-14056.
[27] E. Ma, J. Ding, Tailoring structural inhomogeneities in metallic glasses to enable tensile ductility at room temperature, Materials Today 19(10) (2016) 568-579.
[28] X. Mu, M.R. Chellali, E. Boltynjuk, D. Gunderov, R.Z. Valiev, H. Hahn, C. Kübel, Y. Ivanisenko, L. Velasco, Unveiling the Local Atomic Arrangements in the Shear Band Regions of Metallic Glass, Advanced Materials 33(12) (2021) 2007267.
[29] S. Plimpton, Fast parallel algorithms for short-range molecular dynamics, Journal of Computational Physics 117(1) (1995) 1-19.
[30] M. Mendelev, Y. Sun, F. Zhang, C.-Z. Wang, K.-M. Ho, Development of a semi-empirical potential suitable for molecular dynamics simulation of vitrification in Cu-Zr alloys, The Journal of Chemical Physics 151(21) (2019) 214502.
[31] Z. Pan, T.J. Rupert, Damage nucleation from repeated dislocation absorption at a grain boundary, Computational Materials Science 93 (2014) 206-209.
[32] J. Finney, Modelling the structures of amorphous metals and alloys, Nature 266(5600) (1977) 309-314.
[33] V. Borodin, Local atomic arrangements in polytetrahedral materials, Philosophical Magazine A 79(8) (1999) 1887-1907.
[34] A. Stukowski, Visualization and analysis of atomistic simulation data with OVITO–the Open Visualization Tool, Modelling and Simulation in Materials Science and Engineering 18(1) (2009) 015012.
[35] M. Kumar, E. Nicholson, D.W. Kirk, S.J. Thorpe, C.V. Singh, Short-range structural origins of serration events in metallic glasses, Journal of Alloys and Compounds 787 (2019) 840-850.
[36] T. Lei, J. Shin, D.S. Gianola, T.J. Rupert, Bulk nanocrystalline Al alloys with hierarchical reinforcement structures via grain boundary segregation and complexion formation, Acta Materialia 221 (2021) 117394.
[37] H.-L. Chen, S.-P. Ju, T.-Y. Wu, S.-H. Liu, H.-T. Chen, Investigation of the mechanical properties and local structural evolution of Ti 60 Zr 10 Ta 15 Si 15 bulk metallic glass during tensile deformation: a molecular dynamics study, RSC Advances 5(68) (2015) 55383-55395.
[38] J. Ding, Y.-Q. Cheng, H. Sheng, M. Asta, R.O. Ritchie, E. Ma, Universal structural parameter to quantitatively predict metallic glass properties, Nature Communications 7(1) (2016) 1-10.

[39] W.-P. Wu, D. Şopu, J. Eckert, Molecular Dynamics Study of the Nanoindentation Behavior of Cu64Zr36/Cu Amorphous/Crystalline Nanolaminate Composites, Materials 14(11) (2021) 2756.
[40] S. Yuan, P.S. Branicio, Atomistic simulations of nanoindentation on nanoglasses: Effects of grain size and gradient microstructure on the mechanical properties, Intermetallics 153 (2023) 107782.
[41] M.L. Falk, J.S. Langer, Dynamics of viscoplastic deformation in amorphous solids, Physical Review E 57(6) (1998) 7192-7205.
[42] M. Mendelev, D. Sordelet, M. Kramer, Using atomistic computer simulations to analyze x-ray diffraction data from metallic glasses, Journal of Applied Physics 102(4) (2007) 043501.
[43] K.E. Avila, S. Küchemann, I. Alabd Alhafez, H.M. Urbassek, Shear-transformation zone activation during loading and unloading in nanoindentation of metallic glasses, Materials 12(9) (2019) 1477.
[44] C. Zhong, H. Zhang, Q. Cao, X. Wang, D. Zhang, U. Ramamurty, J. Jiang, Size distribution of shear transformation zones and their evolution towards the formation of shear bands in metallic glasses, Journal of Non-Crystalline Solids 445 (2016) 61-68.
[45] A. Greer, Y. Cheng, E. Ma, Shear bands in metallic glasses, Materials Science and Engineering: R: Reports 74(4) (2013) 71-132.
[46] T. Phan, J. Rigelesaiyin, Y. Chen, A. Bastawros, L. Xiong, Metallic glass instability induced by the continuous dislocation absorption at an amorphous/crystalline interface, Acta Materialia 189 (2020) 10-24.
[47] M.H. Cohen, G. Grest, Liquid-glass transition, a free-volume approach, Physical Review B 20(3) (1979) 1077.
[48] W.L. Johnson, M.D. Demetriou, J.S. Harmon, M.L. Lind, K. Samwer, Rheology and ultrasonic properties of metallic glass-forming liquids: A potential energy landscape perspective, MRS bulletin 32(8) (2007) 644-650.
[49] P.M. Piaggi, M. Parrinello, Entropy based fingerprint for local crystalline order, The Journal of chemical physics 147(11) (2017) 114112.
[50] T. Egami, Atomic level stresses, Progress in Materials Science 56(6) (2011) 637-653.
[51] F. Spaepen, A microscopic mechanism for steady state inhomogeneous flow in metallic glasses, Acta metallurgica 25(4) (1977) 407-415.
[52] M.H. Cohen, D. Turnbull, Molecular transport in liquids and glasses, The Journal of Chemical Physics 31(5) (1959) 1164-1169.
[53] Z. Fan, J. Ding, E. Ma, Making glassy solids ductile at room temperature by imparting flexibility into their amorphous structure, Materials Research Letters 6(10) (2018) 570-583.
[54] Z. Fan, J. Ding, Q.-J. Li, E. Ma, Correlating the properties of amorphous silicon with its flexibility volume, Physical Review B 95(14) (2017) 144211.
[55] L. Wang, Y. Zhang, Z. Zeng, H. Zhou, J. He, P. Liu, M. Chen, J. Han, D.J. Srolovitz, J. Teng, Tracking the sliding of grain boundaries at the atomic scale, Science 375(6586) (2022) 1261-1265.

# Supplemental Materials

**Damage nucleation at ordered grain boundaries**

The damage tolerance of ordered grain boundary (OGB) versions of both grain boundary orientations was examined. Two pure Cu bicrystal samples with ordered grain boundaries, OGB 1 and OGB 2, with exactly the same grain orientations as Complexion 1 and Complexion 2 were created using molecular dynamics (MD) simulations. The OGB samples were relaxed at 10 K using a Nose-Hoover thermo/barostat at zero pressure followed by shear deformation under unrelaxed tensile strain, following the exact same procedure as the amorphous complexion samples in the main text. The dislocations nucleated from the dislocation source were identical in both the OGB samples and the amorphous complexion samples as the center grain (Grain A), the dislocation source, and the simulation cell deformation are the same for all the samples. Both of the OGBs absorbed one partial dislocation (i.e., 0.5 dislocations, using the terminiology introduced in the main text) before damage nucleation at a shear strain of 4.3% in the OGB 1 sample (Fig. S1(a)) and a shear strain of 4.5% in the OGB 2 sample (Fig. S1(b)). Thus, the two ordered grain boundaries have the same dislocation absorption capacity and the difference in the damage tolerance capacity observed for the amorphous complexions cannot be attributed to the misorientation angle between the grains, but rather is directly connected to differences in their local structural order.

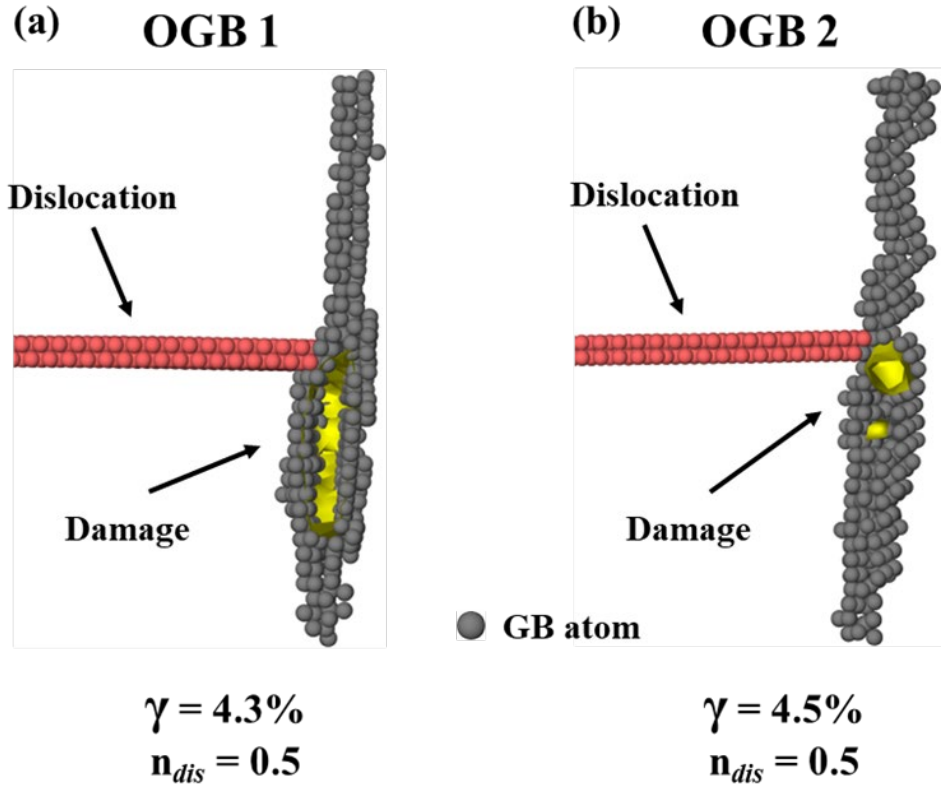

**Fig. S1.** The point of damage nucleation for ordered grain boundary samples (a) OGB 1 and (b) OGB 2, with the shear strain and the number of dislocations absorbed at damage nucleation noted.

**Dislocation motion across the simulation cell**

Fig. S2 shows the Complexion 2 sample during different stages of shear deformation at an applied strain rate of $10^8$ s$^{-1}$. The leading partial dislocations with stacking faults behind them are emitted first from the dislocation source followed by the emission of the trailing partial dislocations. At an applied shear strain of 1.6%, the first leading partial dislocations have been produced by the artificial source and beginning to propagate to the left and right towards the grain boundary complexions (Fig. S2(a)). At a shear strain of 2.2%, the leading partial dislocations interact with the two interfaces and the trailing partial dislocations glide towards the complexions as the applied strain is further increased (Fig. S2(b)). The first trailing partial dislocation is almost absorbed by the complexions as the shear strain is increased to 5.5% and the second dislocation pair is also nucleated in the samples, as shown in Fig. S2(c). As the shear strain is increased to 7.1%, the second trailing dislocation is almost absorbed at both the amorphous complexions with the third pair of dislocation partials being nucleated from the artificial source (Fig. S2(d)).

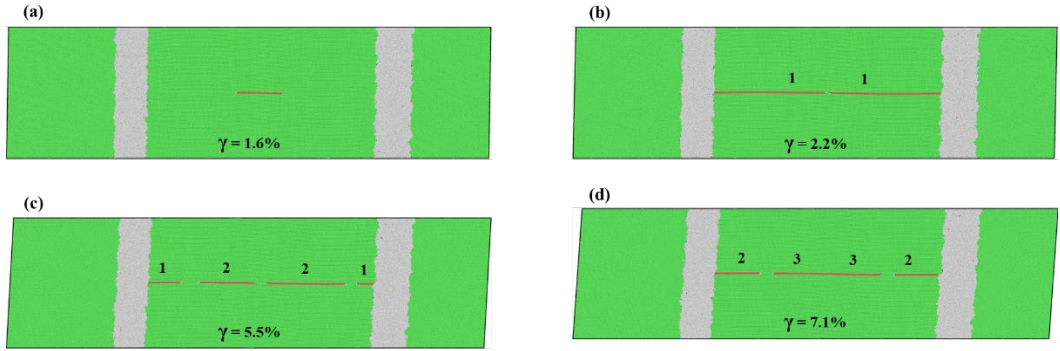

**Fig. S2.** Dislocation emission and absorption observed during shear deformation of the Complexion 2 sample at a shear strain of (a) 1.6%, (b) 2.2%, (c) 5.5%, and (d) 7.1%. The natural numbers label the sequence of dislocations generated from the dislocation source in the center of the specimen. Atoms are colored according to their crystal structure based on Common Neighbor Analysis in OVITO.

**Analysis of dislocation properties**

The dislocations being nucleated in the complexion samples are identical since the middle grain in both of the complexion samples is exactly the same with its X-axis oriented along the [110] direction (slip direction, after passage of a full dislocation or a pair of partial dislocations), the Y-axis oriented along [1$\bar{1}$1] direction (slip plane normal), and the Z-axis oriented along the [1$\bar{1}\bar{2}$] direction (dislocation line length). Furthermore, the pre-tensile strain, the speed of dislocation source displacement, and the shear deformation applied to facilitate dislocation motion and damage nucleation are all the same. This approach was employed to ensure that all the characteristics of the dislocation structure and dislocation motion in both the complexion samples are identical. Analysis of the dislocation character using the dislocation extraction algorithm (DXA) in OVITO [1] shows that the dislocations nucleated from the dislocation source are always Shockley partial dislocations with the same character (Burgers vector of a/6<112>). Fig. S3 shows the leading Shockley partial in Grain A before interaction with the grain boundary complexions, with all FCC atoms removed for clarity. The dislocation velocity is 800 – 850 m/sec before arriving at the complexions and is also identical in both the complexion samples.

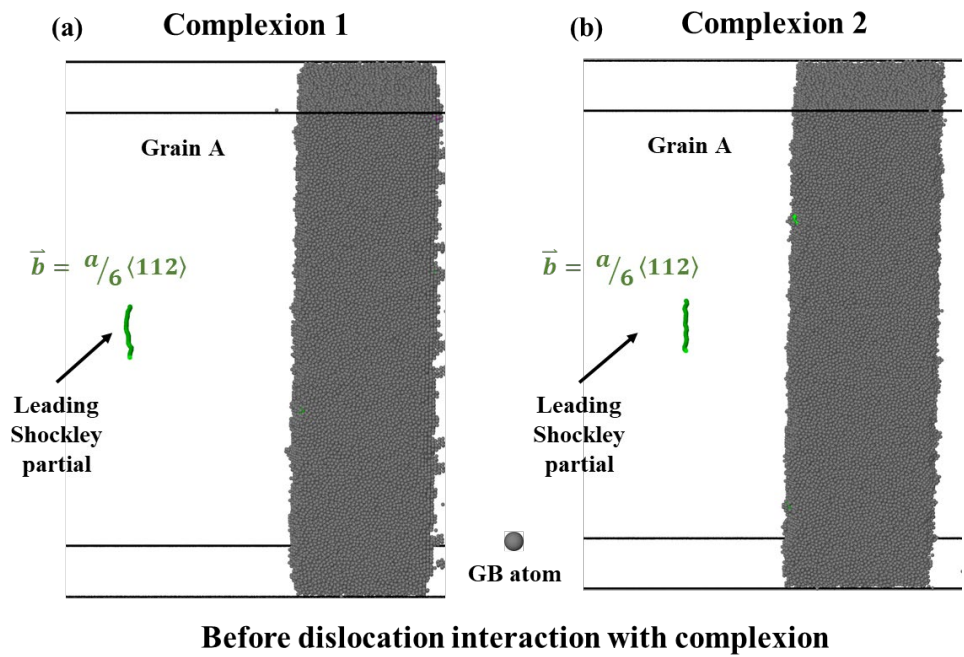

Fig. S3. The leading Shockley partial in the middle Grain A before interaction with (a) Complexion 1 and (b) Complexion 2. All FCC atoms have been removed for clarity.

**Dislocation nucleation in neighboring grains**

The dislocation absorption at amorphous complexions leads to activation of STZs and, upon further increasing the shear strain, can also lead to nucleation of dislocations in the neighboring grains. Fig. S4 shows the complexion samples during deformation, with atoms colored according to their non-affine squared displacement (NASD). Dislocation nucleation into the neighboring grain is observed for Complexion 2, as shown by the green atoms (marked by red arrows) in Grain C, because this sample can be strained to a larger extent.

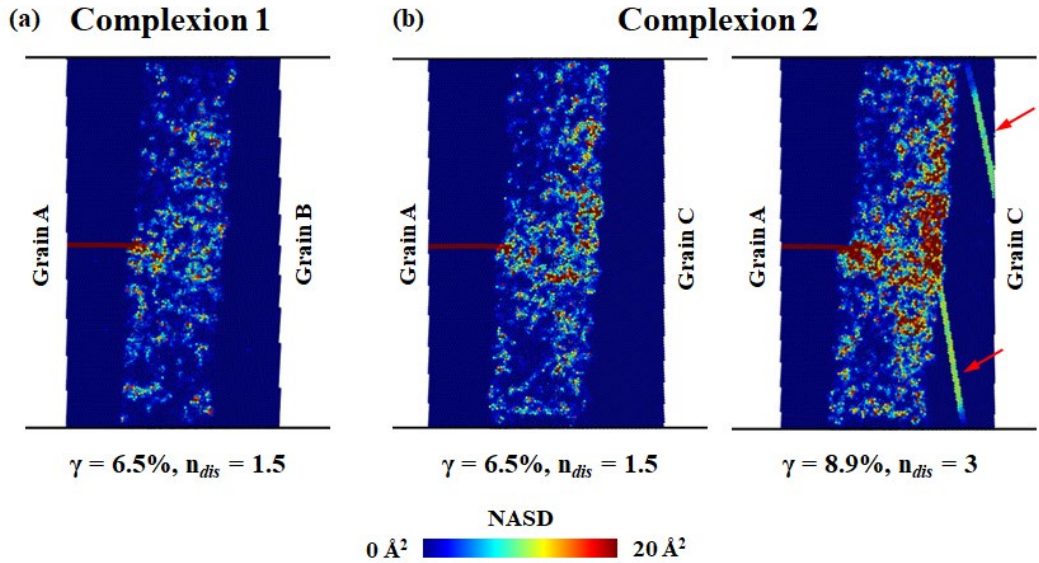

**Fig. S4. The distribution of NASD during deformation in (a) Complexion 1 and (b) Complexion 2. The more damage tolerant Complexion 2 has plastic flow in the outer grains at high strains, as shown by the existence of dislocations (marked by red arrows).**

**Amorphous complexions on left side of the simulation cells**

Each of the complexion samples has two amorphous complexions that are surrounded by the same grains and have identical local structural order. Both of the complexions were for damage tolerance and similar STZ distribution trends were observed for the two identical complexions in each of the samples, so the main text focuses on describing what occurs on the right side of the simulation cell. For completeness, the results for the complexions on the left side are shown here. Fig. S5 shows snapshots of the NASD distribution and the spatial distribution of STZ density in the leftmost complexions of the two samples at the point of damage nucleation. Again,

Complexion 2 is twice as damage tolerant as Complexion 1, identical to the response discussed for the complexion on the right side.

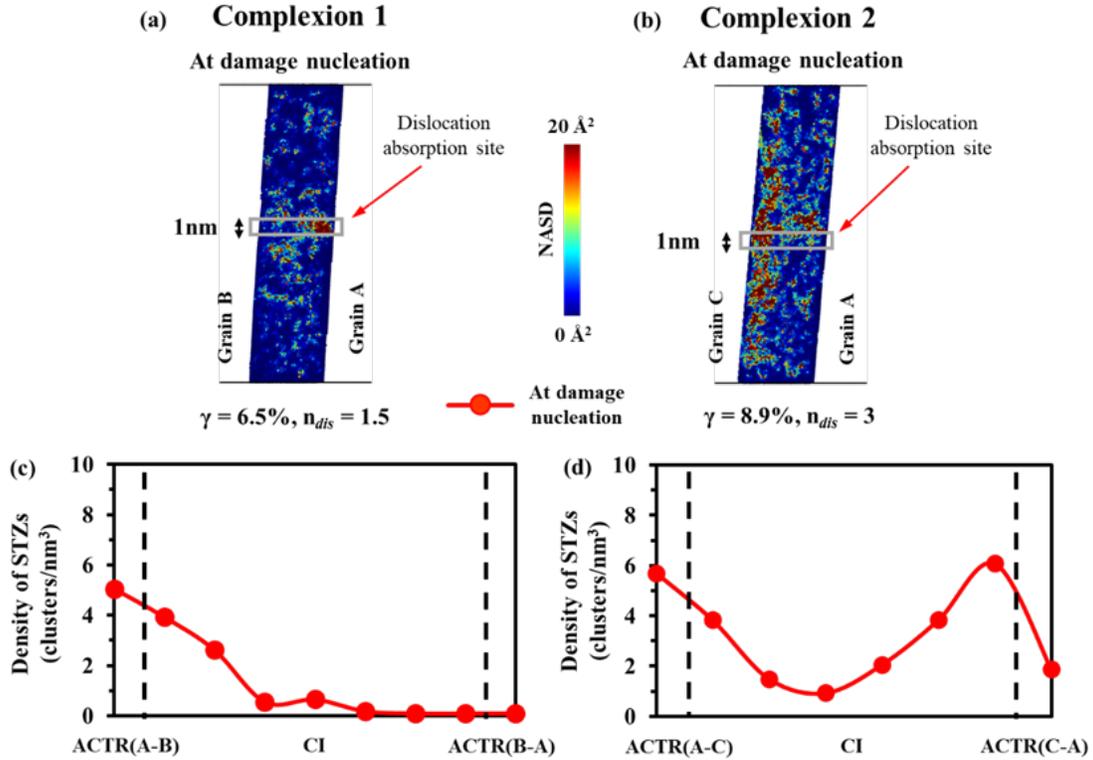

**Fig. S5: The distribution of NASD at damage nucleation in the complexionc on the left side of the simulation cell for (a) Complexion 1 and (b) Complexion 2. Atoms with NASD > 20 Å$^2$ have deformed plastically through STZ operation, with the density of STZs quantified within a 1 nm thick strip normal to the dislocation absorption site. The spatial distributions of STZ density are plotted for (c) Complexion 1 and (d) Complexion 2 with the edges of ACTRs marked by the dashed black lines.**

**References**

[1] A. Stukowski, Visualization and analysis of atomistic simulation data with OVITO–the Open Visualization Tool, Modelling and Simulation in Materials Science and Engineering 18(1) (2009) 015012.